# Modeling single-phase permeability in uniform grain packs

Behzad Ghanbarian

Porous Media Research Lab, Department of Geology, Kansas State University, Manhattan KS 66506 USA (Email address: ghanbarian@ksu.edu)

**Abstract**

Accurate estimation of single-phase permeability ($k$) has broad application in numerous areas, particularly modeling flow and transport in porous materials. Various techniques have been proposed in the literature to estimate $k$ from other medium's properties, such as porosity, grain and/or pore size distribution, and pore connectivity. Among them critical path analysis (CPA) from statistical physics – first developed to model fluid flow in media with broad conductance distributions – has been successfully applied to *heterogeneous* soils and rocks. However, its application to uniform sphere and/or glass bead packs that represent homogeneous porous media with narrow conductance distributions needs to be investigated. In this study, we invoke concepts from CPA and estimate $k$ from average grain diameter and formation factor in uniform sand and glass bead packings. By comparing theory with eight datasets including 105 packs from the literature, we demonstrate that CPA estimates permeability in homogeneous media accurately. We also compare our CPA-based model estimations with those from the Kozeny-Carman, Revil and Cathles, and RGPZ models. Results indicate that the CPA approach estimates $k$ more precisely than other three models studied here.

## 1. Introduction

Modeling fluid flow and transport in fully-saturated porous media requires the knowledge of permeability (*k*), the capability of a medium to allow the passage of fluid through it. Accordingly, the estimation of *k* has been the subject of active research over the past several decades. Various approaches such as bundle of capillary tubes (Burdine, 1953; Purcell, 1949; Xu and Yu, 2008), effective-medium approximations (Doyen, 1988; Ghanbarian and Javadpour, 2017; Richesson and Sahimi, 2019; Sahimi et al., 1983), critical path analysis (Hunt, 2001; Hunt and Sahimi, 2017; Katz and Thompson, 1986) and pore-network models (Bryant and Blunt, 1992; de Vries et al., 2017; Knackstedt et al., 2001) have been applied to determine *k* in porous media. For that purpose, other porous medium's characteristics including size distribution of pores or grains, porosity, specific surface area, formation factor, pore connectivity, and μ-CT images have been typically used. In what follows, we briefly review several *theoretical* approaches used in the literature to model *k* in porous materials. The interested reader is referred to articles by Wen and Gómez-Hernández (1996), Renard and de Marsily (1997), and Sanchez-Vila et al. (2006) for further reviews.

### 1.1. Bundle of capillary tubes model

One of the pioneer methods used to model flow and transport in porous media is based on the concepts of bundle of capillary tubes. In this approach, the complex structure of pore space is replaced with capillary tubes of various sizes. Initial models assumed that capillary tubes were straight (Purcell, 1949). Years later, however, tortuous tubes were bundled to represent irregular pore space in porous media (Burdine, 1953).

Permeability in a bundle of tortuous tubes of the same radius $R$ is given by $k = \phi R^2/(8\tau)$ in which $\phi$ is the porosity and $\tau = (L_e/L_s)^2$ is the tortuosity, the ratio of effective flow path length to sample length (Blunt, 2017). By combining this model with the concept of a hydraulic radius and assuming uniform spheres, Lake et al. (2014) derived the Kozeny-Carman equation (Carman, 1937; Kozeny, 1927), which is typically presented as

$$k = \frac{\bar{D}^2\phi^3}{72\tau(1-\phi)^2} = \frac{\bar{D}^2\phi^3}{180(1-\phi)^2} \qquad (1)$$

where $\bar{D}$ is the representative grain diameter. The numerical prefactor 180 is the product of a constant coefficient 72 and an average tortuosity 2.5 (Lake et al., 2014). The Kozeny-Carman model is known to be valid in uniform glass bead packs with narrow particle/pore size distribution (Bryant et al., 1993). Dullien (1992) stated that, “The Carman-Kozeny equation is of approximate validity. It has been found particularly useful for measuring surface areas of some powders. In the case of particles that deviate strongly from spherical shape, broad particle size distributions, and consolidated media, the Carman-Kozeny equation is often not valid, and therefore, it should always be applied with great caution.”

Eq. (1) was modified by various researchers (e.g., Koltermann and Gorelick, 1995; Kuang et al., 2011; Panda and Lake, 1994; Porter et al., 2013) to estimate $k$ for more complex media. For example, Van Der Marck (1996) and Mavko and Nur (1997) incorporated the effect of percolation threshold (or critical porosity) in Eq. (1). However, the main drawback of the bundle of capillary tubes approach is neglecting the effect of pore connectivity (Hunt et al., 2014; Sahimi, 2011). Nonetheless, it has been widely combined with the power-law pore size distribution concept to model $k$ in fractal porous media (Guarracino et al., 2014; Wang et al., 2011; Wei et al., 2018; Xu and Yu, 2008; Yu and Cheng, 2002).

### 1.2. Johnson et al. (1986) model

Johnson, Koplik and Schwartz (Johnson et al., 1986) introduced a characteristic pore size ($\Lambda$), a measure of the size of the dynamically connected pores that is independent of dead-end and isolated pores (Banavar and Schwartz, 1987). In the Johnson et al. (1986) model, $\Lambda$ is related to permeability via

$$k = \frac{\Lambda^2}{8F}, \quad (2)$$

in which $F$ is the formation factor ($= \sigma_f / \sigma_b$ in which $\sigma_f$ and $\sigma_b$ represent the electrical conductivity of fluid and bulk, respectively).

Banavar and Johnson (1987) proposed theoretical relationships to determine $\Lambda$ from the inflection point on the mercury porosimetry curve (i.e., $\Lambda = a l_{inf}$ in which $a$ is a constant and $l_{inf}$ is the pore size corresponding to the inflection point). Those authors assumed that pores were cylindrical and considered two pore geometries: (1) pore length was constant and (2) pore length was equal to its diameter. For the former, they found $a$ = 0.25, while for the latter $a$ = 0.167. If one assumes that a porous medium is probably a mixture of the two pore geometries considered by Banavar and Johnson (1987), one may use an average value $a$ = 0.21, which is close to the value 0.19 that Revil et al. (2014) applied.

In another study, Revil and Cathles (1999) proposed $\Lambda = \bar{D}/m(F-1)$ to link $\Lambda$ to the average grain diameter. If this expression is combined with Eq. (2), one has

$$k = \frac{(\bar{D})^2}{32m^2F(F-1)^2}, \quad (3)$$

where $m$ is the cementation exponent in Archie's law i.e., $F = \phi^{-m}$ (Archie, 1942). Although Eq. (3) was frequently attributed to Revil and Cathles (1999), those authors did not presented their permeability model in the form of Eq. (3).

Using a terminology similar to Revil and Cathles (1999), Glover et al. (2006) proposed (known as the RGPZ model):

$$k = \frac{(\bar{D})^2 \phi^{3m}}{4am^2} \quad (4)$$

where $a$ = 8/3 for quasi-spherical grains. Glover et al. (2006) recommended that $\bar{D}$ in Eq. (4) should be determined from the geometric mean of grain diameters. Using experiments from 8 glass bead packs, Glover et al. (2006) showed that their model estimated $k$ more accurately than the Kozeny-Carman equation.

### 1.3. Critical path analysis (CPA)

Based on arguments by Ambegaokar et al. (1971) and Pollak (1972), fluid flow in a disordered porous medium with broad distribution of conductances, $f(g)$, is controlled by conductances whose magnitudes are greater than some critical conductance $g_c$, which is defined as the smallest conductance required to form a conducting sample-spanning cluster. Within the framework of CPA, permeability in a network of pores is dominated by highly-conducting pores, while low-conducting ones have trivial contribution to the effective permeability (Hunt et al., 2014; Sahimi, 2011).

Katz and Thompson (1986) were the first to apply concepts from critical path analysis to estimate permeability of rocks from the critical pore diameter and formation factor. They presumed that pore diameter was linearly proportional to its length and proposed

$$k = \frac{d_c^2}{cF} \quad (5)$$

in which $d_c$ is the critical pore diameter and $F$ is the formation factor. Katz and Thompson (1986) indicated that the critical pore diameter can be well determined from the mode of the pore size distribution, corresponding to the inflection point on the mercury intrusion porosimetry curve. In Eq. (5), $c$ is a constant coefficient equal to 226 (Katz and Thompson, 1986). However, other $c$ values were later proposed for various circumstances. For a recent review, see Table 1 in Ghanbarian et al. (2016b).

Using several hundred rock samples including sandstones and carbonates with measured permeability spanned near 8 orders of magnitude (from $10^{-3}$ to $10^{5}$ mD), Thompson (1991) showed that Eq. (5) with $c$ = 226 estimated permeability accurately in sandstones, carbonates, and meta-morphic rocks. He found agreement between theory and experiment within a factor of two except for samples with experimental issues associated with either measurement errors, heterogeneities, or cracks. Similarly, Ghanbarian et al. (2016b) demonstrated that CPA estimated $k$ within a factor of two of measurements in tight-gas sandstones. However, they used $c$ = 53.5, in accord with Skaggs (2011). In another study, Daigle (2016) used CPA with $c$ = 32 to determine $k$ in a variety of rock types including sandstone, carbonate, and clay-rich samples. He assumed that pore size distribution followed power-law behavior, formation factor conformed to universal scaling from percolation theory, and demonstrated that the CPA model estimates matched measured values well.

### 1.4. Objectives

Critical path analysis has been successfully applied to estimate permeability in soils and rocks, which typically have broad pore size distributions (Ghanbarian et al., 2017; Hunt et al., 2014; Hunt, 2001; Katz and Thompson, 1986). However, its application for estimating permeability in uniform sand and glass bead packs, representing homogeneous media with relatively narrow conductance distribution, has not been addressed yet, particularly using a database including 105 experiments. Therefore, the main objectives of this study are to: (1) invoke CPA for estimating permeability $k$ in the packings of uniform grains, and (2) compare CPA permeability estimates with predictions from the Kozeny-Carman, Revil and Cathles (1999), and RGPZ models.

## 2. Materials and Methods

The data used in this study are experimental observations measured in uniform glass bead and sand packs from eight datasets available in the literature. In what follows, we briefly describe each dataset, and the interested reader is referred to the original published articles for further detail.

**- Chauveteau and Zaitoun (1981)**

Chauveteau and Zaitoun (1981) packed glass beads of different diameters to compose porous media with similar pore shapes but different pore sizes (see their Table 1). All seven uniform glass bead packs had similar porosity about 0.4. The flow experiment was performed with a 400 ppm xanthan solution. They reported glass bead diameter, porosity, and permeability. However, formation factor was not measured and, thus, is not available for this dataset.

**- Biella et al. (1983)**

Biella et al. (1983) measured porosity, formation factor, and permeability in one- and two-component clean sand samples. However, in this study only mono-granular media including 11 packs composed of rounded sands are used (see their Table 1). The samples were obtained by washing and sieving alluvial material collected from a quaternary deposit. Those authors used 11 classes and standard sieve sizes between 0.1 and 8 mm. First, all the samples were saturated with water at similar fluid resistivity, viscosity, and temperature. Then, electrical resistivity and permeability were measured simultaneously using a sand-settling cell. More specifically, permeability was determined using the constant head method, and formation factor using four potential electrodes.

**- Moghadasi et al. (2004)**

This dataset includes both glass bead and sand packs. The average porosity was 0.384 in sand and 0.380 in glass bead packs. Distilled water was first pumped through each pack for about an hour. Permeability was then determined using the steady-state method. For that purpose, pressure values were recorded at short time intervals.

**- Glover et al. (2006)**

The uniform packs from Glover et al. (2006) were composed of glass spheres with a high degree of sphericity and a tight tolerance. They were randomly packed into cylinders with 2.54 cm in diameter and 2.5 to 5 cm length. The samples were saturated using an aqueous solution of 0.1 M of sodium chloride (NaCl) of a known density and electrical resistivity, and porosity was determined via the gravimetric method. Permeability was calculated at five flow rates, and an arithmetic average was used to represent each pack. Electrical resistivity was measured using a Solartron 1260 impedance analyzer, and formation factor was calculated from the resistivity at 1 kHz.

**- Glover and Walker (2009)**

Glover and Walker (2009) applied the same procedure used by Glover et al. (2006) to measure porosity, electrical resistivity and permeability in seven glass bead packs. However, as Table 1 shows, Glover and Walker (2009) used glass beads of sizes different from those packed by Glover et al. (2006).

**- Glover and Dery (2010)**

Silica glass beads were washed several times in distilled water and acetone and then packed to measure porosity and permeability. Grain size distribution was determined using optical microscopy, image analysis, and laser diffraction particle size analyzer. The geometric mean dimeter of each set of glass beads was calculated by fitting the log-normal probability density function to the measured grain size distributions (see $\overline{D}$ given in Table 1). Porosity was measured with the porosimetry method using both helium and mercury. However, only the helium porosity values were used in this study. Permeability was determined from the measured flow rate and pressure drop under low Reynolds number and laminar conditions ($R_e < 3.5\times10^{-3}$).

**- Koch et al. (2012)**

Samples from Koch et al. (2012) include non-compacted and compacted quartz sand packs. Koch et al. (2012) compacted samples using continued shaking and refilling the sample container. Total porosity of each pack was calculated from its bulk density by assuming that particle density was 2.65 gr/cm$^3$. The average difference in porosity between the compacted and non-compacted samples was about 7% (see Table 1). Formation factor was determined from the slope of fluid conductivity-bulk conductivity plot. Permeability for each pack was calculated from fluid flow under constant head measurements in a sample of diameter of 5.1 cm and length of 5 cm.

**- Kimura (2018)**

In this dataset including 34 uniform packs, grain diameters were measured using 21 sieves with equal intervals. Grain density of glass beads and sands was determined via pycnometer. Porosity of each uniform pack was calculated from the measured grain and bulk densities. Kimura (2018) used two pairs of electrodes, composed of thin stainless-steel rods, to measure electrical resistivity and determined the value of formation factor from the linear relationship between fluid and bulk electrical conductivities (see his Fig. 1). The constant head method with a $6.8\times10^{-2}$ m water head difference was applied to calculate permeability for each pack.

**- Estimating permeability *k* in uniform glass bead and sand packs**

Estimating $k$ via the Kozeny-Carman, Revil and Cathles, RGPZ, and CPA models requires representative grain diameter determination. In the literature, arithmetic, geometric, and harmonic means (Koltermann and Gorelick, 1995; Porter et al., 2013; Urumović and Urumović Sr, 2016; and references therein) have been used to estimate $\overline{D}$. Since in uniform sphere packs with narrow grain size distributions, the three averages are not greatly different, we used the arithmetic mean for all datasets except Glover and Dery (2010) for which those authors reported the geometric mean.

In contrast to the Kozeny-Carman model estimating permeability from $\overline{D}$ and $\phi$, the Revil and Cathles (1999) and RGPZ models estimate $k$ from $\overline{D}$, $\phi$, and the formation factor, $F$, while the CPA model from $d_c$ and $F$. Since the pore size distribution of the sand and/or glass bead packs studied here are not available, $d_c$ was estimated from the value of $\overline{D}$ and a relationship by Ng et al. (1978). They proposed that for random mono-sized sphere packs with $\phi$ = 0.4 the average pore throat diameter can be approximated by $0.21D$ in which $D$ is the grain diameter. Following Ng et al. (1978), we set $d_c \approx 2\overline{r_t} = 0.42\overline{D}$ in which $\overline{r_t}$ is the average pore throat radius. By comparison with the experiments, we show that this approximation yields accurate permeability estimations in uniform glass bead and/or sand packs. However, the assumption $d_c \approx 2\overline{r_t} = 0.42\overline{D}$ may cause uncertainties in the permeability estimation, particularly for packings with porosities substantially different than 0.4. More specifically, Ng et al. (1978) reported $\overline{r_t} = 0.414D$ for simple cubic and mono-sized sphere packs with $\phi$ = 0.476 indicating that $d_c$ should be a function of pore space structure and grain arrangement.

The measured value of formation factor is not available for the Chauveteau and Zaitoun (1981), Moghadasi et al. (2004) and Glover and Dery (2010) datasets. Accordingly, we approximated $F$ by $\phi^{-1.5}$, derived theoretically by Sen et al. (1981) for mono-sized sphere packs using the self-consistent model.

Recently, Ghanbarian et al. (2016a) proposed a theoretical scaling of Poiseuille's law modified for flow in cylindrical pores with rough surfaces. More specifically, for isotropic systems they derived $g_h \propto r^{2(4-D_S)-\frac{3-D_S}{2D_S-3}}$ in which $g_h$ is the hydraulic conductance, $r$ is the average pore radius, and $D_s$ represents the surface fractal dimension. For sand and glass bead packs with smooth pore-solid interfaces one may approximate pores by cylindrical tubes, presume that $D_s \approx 2$ and, thus, $g_h \propto r^3$. To estimate permeability using CPA and Eq. (5) we accordingly set $c$ = 72.2 (see Table 1 in Ghanbarian et al. (2016b)), as Skaggs (2011) recommended. The assumption $g_h \propto r^3$ is also consistent with water relative permeability estimations in mono-sized sphere packs (Ghanbarian, 2019) and self-similar media in which pore length is proportional to its radius (Hunt, 2001).

**- Model evaluation criterion**

To assess the reliability of various models in this study, the root mean square log-transformed error (RMSLE) is determined as follows:

$$RMSLE = \sqrt{\frac{1}{N}\sum_{i=1}^{N}[\log(k_{est}) - \log(k_{meas})]^2} \quad (6)$$

where $N$ represents the number of samples and $k_{est}$ and $k_{meas}$ are respectively the estimated and measured permeability values.

## 3. Results and Discussion

### 3.1. Models evaluation

In this section, we present the results obtained from comparing the Kozeny-Carman, Revil and Cathles (1999), RGPZ, and CPA models with the experimental measurements and discuss each model's accuracy and reliability. Figure 1 shows permeability estimates by the four models against the measured values spanning near eight orders of magnitude for all 105 uniform packs summarized in Table 1. As can be seen, although the Kozeny-Carman model generally overestimated $k$ (Fig. 1a), the Revil and Cathles (1999) model underestimated the value of permeability (Fig. 1b). The Revil and Cathles (1999) model estimations, however, are more accurate than those by the Kozeny-Carman equation (RMSLE = 0.26 vs. 0.32 $\mu m^2$).

For the sake of comparison, we also estimated $k$ via the Kozeny-Carman equation and the geometric grain diameter. Results, not shown, indicated that when the geometric mean was used the RMSLE value slightly decreased to 0.31 $\mu m^2$. This clearly demonstrates that the choice of arithmetic, geometric, or harmonic mean should not affect permeability estimation in such uniform glass bead and/or sand packs.

Several studies in the literature (Glover et al., 2006; Koch et al., 2012; Mavko and Nur, 1997; Van Der Marck, 1996) showed that the Kozeny-Carman model may overestimate permeability, particularly in media with low porosity. For example, Koch et al. (2012) stated that the Kozeny–Carman model overpredicts permeability due to its inherent overestimation of the fraction of connected porosity. To address this, the following modified form of the Kozeny-Carman equation that includes the effect of percolation threshold (or critical porosity) was proposed:

$$k = \frac{\bar{D}^2(\phi - \phi_c)^3}{180(1-\phi)^2} \tag{7}$$

where $\phi_c$ is the critical porosity. van Der Marck (1996) found $\phi_c$ = 0.03 in mono-sized and bi-disperse sphere packs and demonstrated that Eq. (7) estimated permeability accurately over a wide range of porosity e.g., $0.03 < \phi < 0.4$ (see his Fig. 1). We used Eq. (7) with $\phi_c$ = 0.03 to estimate permeability for 105 packs (results not shown) and found RMSLE = 0.26, which indicates that the permeability estimations were improved compared to the traditional Kozeny-Carman model (Eq. 1).

Figs. 1c and 1d show permeability estimations by the RGPZ and CPA models versus the measured values for the 105 uniform packs. For the former we found RMSLE = 0.19 and for the latter RMSLE = 0.16. These two models estimated permeability around the 1:1 line and more accurately than the Kozeny-Carman and Revil and Cathles (1999) models. However, CPA's estimations are slightly more precise than RGPZ's predictions.

Glover et al. (2006) compared estimated permeability via the RGPZ and Kozeny-Carman models with measured permeability for 65 sandstone and carbonate samples. They showed that the RGPZ model with geometric mean diameter resulted in remarkably more accurate estimations than the Kozeny-Carman model, which mainly overestimated $k$ in such consolidated natural porous media (see their Fig. 3).

The porosity of the sand and glass bead packs studied here varies over a relatively wide range from 0.36 to 0.49 (see Table 1). We should point out that the relationship $d_c \approx 2\overline{r_t} = 0.42\overline{D}$ was deduced for random mono-sized sphere packs with $\phi$ = 0.4 (Ng et al., 1978). As stated earlier, one should not expect that $d_c \approx 2\overline{r_t} = 0.42\overline{D}$ provides accurate estimate of $d_c$ for packings whose porosities differ from 0.4 remarkably. In fact, Ng et al. (1978) reported $\overline{r_t} = 0.414D$ for simple cubic and mono-sized sphere packs with $\phi$ = 0.476. This indicates that, as expected, the value of critical pore diameter depends on grain arrangement and, consequently, pore space structure.

Interestingly, CPA estimated $k$ more accurately than the Kozeny-Carman model (see Fig 1). The former is known to be valid in heterogeneous porous media with broad conductance distributions, while the latter in homogeneous and unconsolidated porous materials with narrow grain size distributions. For example, for the mono-sized sand pack ($\overline{D}$ = 1000 μm) in the Moghadasi et al. (2004) dataset, CPA estimation was 583.7 μm$^2$ (~ 39% greater the actual permeability value), while the Kozeny-Carman model prediction was 838.2 μm$^2$ (~ 99% greater the measured value). Given that the value of formation factor was estimated from $F = \phi^{-1.5}$ in this dataset, both CPA and Kozeny-Carman models estimated $k$ from the same input parameters i.e., arithmetic mean grain diameter ($\overline{D}$) and porosity ($\phi$). This indicates under the same circumstances, CPA can provide more accurate estimations than the Kozeny-Carman equation.

CPA is known to be an appropriate upscaling technique for porous media with broad conductance distributions. Although the term "broad" has not been satisfactorily defined in the literature, results presented in Fig. 1 clearly demonstrate that CPA provides accurate estimates of single-phase permeability in uniform glass bead and sand packs with narrow conductance distributions. Shah and Yortsos (1996) argued that because the exponent in the hydraulic conductance-radius relationship (i.e., $g_h \sim r^4$) is large, even homogenous media might possess a broad conductance distribution and, thus, CPA should be valid in such materials. Accordingly, even for uniform glass bead and sand packs whose pore sizes only span about one order of magnitude or less, one may expect CPA to be reasonably accurate, as experimentally shown in this study.

We should point out that concepts from CPA are also applicable to clay-rich media with non-negligible surface conduction. In such materials, in addition to bulk conduction ($\sigma_b$), surface conduction ($\sigma_s$) may effectively contribute to electrical conductivity (Revil et al., 2014b). To accurately determine the value of formation factor and precisely estimate permeability, one needs to measure electrical conductivity using a highly saline brine so that $\sigma_b \gg \sigma_s$. Although we mainly addressed applications from CPA to sand and glass bead packs, this approach has been successfully used to estimate permeability in natural porous media such as marine mudstones (Daigle, 2016), tight-gas sandstones (Ghanbarian et al., 2016b), shales (Zhang and Scherer, 2012), and soils (Ghanbarian et al., 2017).

### 3.2. Critical pore diameter determination

Although we demonstrated that the relationship $d_c \approx 0.42\bar{D}$ yielded accurate results in uniform glass bead and/or sand packs with porosity near 0.4 (see Fig. 1d), one should expect this approximation to cause uncertainties in the estimation of $k$ for multi-dispersed grain packs. In such porous media, the value of critical pore diameter $d_c$ may be determined from either X-ray images (Arns et al., 2005; Koestel et al., 2018), water retention data (Ghanbarian et al., 2017) or mercury intrusion porosimetry curve (Katz and Thompson, 1986). For example, Arns et al. (2005) investigated relationships used to estimate permeability from pore size properties in Fontainebleau sandstones and three-dimensional computed tomography (CT) images. They considered relationships based on the ratio of pore volume to surface area, critical pore diameter (associated with mercury intrusion porosimetry data), as well as characteristic pore sizes associated with nuclear magnetic resonance relaxation time. Arns et al. (2005) found that all those relationships provided good agreement with their lattice-Boltzmann simulations. However, permeability values estimated based on critical pore diameter (and critical path analysis) were found to be the most reliable (Arns et al., 2005).

### 3.3. Correlation between permeability and characteristic length scale

In Fig. 2, we show the measured permeability $k$ as a function of the average grain diameter $\overline{D}$ and critical pore diameter $d_c$ for 105 packs summarized in Table 1. Strong correlation between $k$ and $\overline{D}$ was previously reported in glass bead packs; see e.g. Beavers et al. (1973) and Bryant et al. (1993). Given that the measured permeability value spans over seven orders of variations, the high correlation coefficient $R^2$ = 0.99 is remarkable, and the exponent 1.92 is only 4 percent less than the theoretical value of 2 in Eqs. (1), (3), (4) and (5). The exponent 2 in those equations has a physical basis—permeability is proportional to some length scale squared—and is also consistent with the results of Thomeer (1960) and Swanson (1981). Thomeer (1960) plotted mercury intrusion volume as a function of capillary pressure on a log-log scale and approximated the resulting curve by a hyperbola.

Following Thomeer (1960), Swanson (1981) determined the capillary pressure corresponding to the apex of the hyperbolic curve ($P_{capex}$). He found that permeability was highly correlated to $P_{capex}$ in a power-law form (i.e., $k \propto [S_{Hg}/P_c]_{apex}^{\alpha}$ in which $S_{Hg}$ and $P_c$ are respectively mercury saturation and capillary pressure corresponding to the apex) with $\alpha \approx 2$ for clean sandstones and carbonates.

We should also point out that techniques used to measure porosity, formation factor and permeability were different from one dataset to another. Accordingly, one may expect slight scatter in the $k$-$\bar{D}$ data due to various measurement precisions. However, such an effect should be trivial since $R^2$ = 0.99 and permeability values span more than seven orders of magnitude (Fig. 2).

**3.4. Formation factor in uniform packings**

Figure 3 shows the measured formation factor $F$ against the measured porosity $\phi$ for 78 uniform packings of sand and glass bead on the log-log scale. We also show the theoretical predictions by $F = \phi^{-1.5}$, developed for mono-sized sphere packs using the self-consistent model (Sen et al., 1981). As can be seen in Fig. 3, although the data follow the theoretical predictions, they are scattered. This might be because the formation factor was measured in uniform glass bead and sand packs rather than perfectly mono-sized grain packs.

Recently, Ghanbarian et al. (2013) combined concepts from percolation theory and finite-size scaling and proposed a geometric model for tortuosity (their Eq. 8). If we replace $\theta$ (water content) with $\phi$ (porosity), set $C/L_s$ and $\theta_t$ equal to zero and use $\nu$ = 0.88 and $D_{opt}$ = 1.43 (universal values from percolation theory in three dimensions) in the Ghanbarian et al. (2013) model, one has $\tau_g = \phi^{-0.378}$. Assuming that geometric and electrical tortuosities (respectively $\tau_g$ and $\tau_e$) are similar in media with narrow pore/particle size distributions (Ghanbarian et al., 2013a), combining $\tau_g = \phi^{-0.378}$ with $\tau_e = \phi F$ gives $F = \phi^{-1.378}$. Interestingly, the exponent -1.378 is not greatly different from -1.5 proposed theoretically by Sen et al. (1981). This indicates both percolation theory and self-consistent models provide consistent results for tortuosity and consequently formation factor in uniform and homogeneous systems.

## 4. Conclusions

Critical path analysis (CPA) is known to be valid in heterogeneous porous media with broad conductance distributions. However, the term ‘‘broad’’ has not satisfactorily been quantified in the literature. In this study, we examined whether CPA can accurately estimate the single-phase permeability *k* in homogeneous porous media. By comparison with 105 samples from the literature, we showed that the proposed CPA-based model estimated *k* from the average grain diameter and formation factor in uniform glass bead and sand packs precisely. The accuracy of the proposed model was also compared to that of the Kozeny-Carman equation, which is valid in unconsolidated porous materials with narrow grain size distribution, as well as the Revil and Cathles (1999) and RGPZ (Glover et al., 2006) models. Results indicated that the CPA-based model estimated *k* more accurately than the other three models studied here.

## Acknowledgment

The author acknowledges the Associate Editor, WeiCheng Lo, and two anonymous reviewers for constructive and fruitful comments that improved the quality of this manuscript. The author is also grateful to Kansas State University for supports through faculty startup funds.

Table 1. Properties of the uniform glass bead and sand packs studied here.

| **Reference** | **Pack** | **Grain diameter range (μm)** | $\bar{D}$ **(μm)** | $\phi$ | **m** | $F$ | **Measured *k* (μm2)** | $d_c$ **(μm)** |
|---|---|---|---|---|---|---|---|---|
| Chauveteau & Zaitoun (1981) | Glass bead | 400-500 | 450 | 0.40 | 1.5 | 3.95 | 137 | 189.0 |
| | Glass bead | 200-250 | 225 | 0.40 | 1.5 | 3.95 | 36 | 94.5 |
| | Glass bead | 80-100 | 90 | 0.40 | 1.5 | 3.95 | 8.4 | 37.8 |
| | Glass bead | 40-50 | 45 | 0.41 | 1.5 | 3.81 | 2.4 | 18.9 |
| | Glass bead | 20-30 | 25 | 0.41 | 1.5 | 3.81 | 0.66 | 10.5 |
| | Glass bead | 10-20 | 15 | 0.41 | 1.5 | 3.81 | 0.21 | 6.3 |
| | Glass bead | 8-15 | 11.5 | 0.41 | 1.5 | 3.81 | 0.11 | 4.8 |
| Biella et al. (1983) | Sand | 100-200 | 150 | 0.45 | - | 3.92 | 6.7 | 63.0 |
| | Sand | 200-400 | 300 | 0.43 | - | 4.10 | 49.2 | 126.0 |
| | Sand | 400-600 | 500 | 0.40 | - | 4.05 | 107.7 | 210.0 |
| | Sand | 600-1000 | 800 | 0.41 | - | 4.29 | 205.1 | 336.0 |
| | Sand | 1000-1600 | 1300 | 0.40 | - | 4.20 | 810.2 | 546.0 |
| | Sand | 1600-2000 | 1800 | 0.39 | - | 4.31 | 1261.4 | 756.0 |
| | Sand | 2000-3150 | 2575 | 0.37 | - | 4.77 | 2563.8 | 1081.5 |
| | Sand | 3150-4000 | 3575 | 0.38 | - | 4.88 | 5127.6 | 1501.5 |
| | Sand | 4000-5000 | 4500 | 0.37 | - | 4.64 | 5640.4 | 1890.0 |
| | Sand | 5000-6300 | 5650 | 0.37 | - | 4.70 | 8204.2 | 2373.0 |
| | Sand | 6300-8000 | 7150 | 0.37 | - | 4.70 | 12306.3 | 3003.0 |
| Moghadasi et al. (2004) | Sand | 180-250 | 192 | 0.383 | 1.5 | 4.22 | 21.4 | 80.6 |
| | Sand | 250-425 | 265 | 0.383 | 1.5 | 4.22 | 60.3 | 111.3 |
| | Sand | 400-500 | 410 | 0.384 | 1.5 | 4.20 | 121 | 172.2 |
| | Sand | 1000 | 1000 | 0.385 | 1.5 | 4.19 | 727 | 420.0 |
| | Glass bead | 180-300 | 245 | 0.379 | 1.5 | 4.29 | 44.1 | 102.9 |
| | Glass bead | 250-425 | 338 | 0.379 | 1.5 | 4.29 | 78.9 | 142.0 |
| | Glass bead | 400-600 | 480 | 0.380 | 1.5 | 4.27 | 159 | 201.6 |
| | Glass bead | 1000 | 1000 | 0.383 | 1.5 | 4.22 | 705 | 420.0 |
| Glover et al. (2006) | Glass bead | 20±0.5 | 20 | 0.401 | - | 3.90 | 0.24 | 8.4 |
| | Glass bead | 45±1.2 | 45 | 0.391 | - | 4.02 | 1.6 | 18.9 |
| | Glass bead | 106±4 | 106 | 0.394 | - | 4.05 | 8.12 | 44.5 |
| | Glass bead | 250±15 | 250 | 0.398 | - | 3.98 | 50.5 | 105 |
| | Glass bead | 500±25 | 500 | 0.381 | - | 4.09 | 186.8 | 210 |
| | Glass bead | 1000±34 | 1000 | 0.395 | - | 3.91 | 709.9 | 420 |
| | Glass bead | 2000±67 | 2000 | 0.386 | - | 4.14 | 2277.3 | 840 |
| | Glass bead | 3350±184 | 3350 | 0.397 | - | 3.93 | 7707.0 | 1407 |
| Glover & Walker (2009) | Glass bead | 3000±154 | 3000 | 0.398 | - | 4.21 | 4892 | 1260 |
| | Glass bead | 4000±198 | 4000 | 0.385 | - | 4.38 | 6706 | 1680 |
| | Glass bead | 5000±267 | 5000 | 0.376 | - | 4.65 | 8584 | 2100 |
| | Glass bead | 6000±255 | 6000 | 0.357 | - | 5.31 | 8262 | 2520 |
| | Glass bead | 256±44 | 256 | 0.399 | - | 4.01 | 41.2 | 107.5 |
| | Glass bead | 512±88 | 512 | 0.389 | - | 4.36 | 164 | 215.0 |
| | Glass bead | 181±31 | 181 | 0.382 | - | 4.39 | 18.6 | 76.0 |
| Glover & Dery (2010) | Quartz glass bead | - | 1.05 | 0.411 | 1.5 | 3.80 | 0.00057 | 0.441 |
| | Quartz glass bead | - | 2.11 | 0.398 | 1.5 | 3.98 | 0.00345 | 0.886 |
| | Quartz glass bead | - | 5.01 | 0.380 | 1.5 | 4.27 | 0.0181 | 2.104 |
| | Quartz glass bead | - | 11.2 | 0.401 | 1.5 | 3.94 | 0.0361 | 4.704 |
| | Quartz glass bead | - | 21.5 | 0.383 | 1.5 | 4.22 | 0.228 | 9.03 |
| | Quartz glass bead | - | 31 | 0.392 | 1.5 | 4.07 | 0.895 | 13.02 |
| | Quartz glass bead | - | 47.5 | 0.403 | 1.5 | 3.91 | 1.258 | 19.95 |
| | Quartz glass bead | - | 104 | 0.394 | 1.5 | 4.04 | 6.028 | 43.68 |
| | Quartz glass bead | - | 181 | 0.396 | 1.5 | 4.01 | 21.53 | 76.02 |
| | Quartz glass bead | - | 252 | 0.414 | 1.5 | 3.75 | 40.19 | 105.84 |
| | Quartz glass bead | - | 494 | 0.379 | 1.5 | 4.29 | 224 | 207.48 |
| | Quartz glass bead | - | 990 | 0.385 | 1.5 | 4.19 | 866.7 | 415.8 |
| Koch et al. (2012) | Quartz sand | - | 180 | 0.47 | - | 3.77 | 17.6 | 75.6 |
| | Quartz sand | - | 270 | 0.45 | - | 3.55 | 53.1 | 113.4 |
| | Quartz sand | - | 660 | 0.47 | - | 3.25 | 129 | 277.2 |
| | Quartz sand | - | 180 | 0.48 | - | 3.14 | 20.8 | 75.6 |
| | Quartz sand | - | 230 | 0.49 | - | 3.40 | 33 | 96.6 |
| | Quartz sand | - | 320 | 0.49 | - | 3.26 | 67.5 | 134.4 |
| | Quartz sand | - | 500 | 0.49 | - | 3.12 | 171 | 210 |
| | Quartz sand | - | 680 | 0.48 | - | 3.10 | 280 | 285.6 |
| | Quartz sand | - | 860 | 0.49 | - | 3.34 | 394 | 361.2 |
| | Quartz sand | - | 180 | 0.39 | - | 4.12 | 11.1 | 75.6 |
| | Quartz sand | - | 270 | 0.39 | - | 3.75 | 24 | 113.4 |
| | Quartz sand | - | 660 | 0.41 | - | 3.97 | 75 | 277.2 |
| | Quartz sand | - | 180 | 0.40 | - | 3.23 | 11.7 | 75.6 |
| | Quartz sand | - | 230 | 0.40 | - | 3.55 | 19.8 | 96.6 |
| | Quartz sand | - | 320 | 0.42 | - | 3.64 | 38.1 | 134.4 |
| | Quartz sand | - | 500 | 0.42 | - | 3.52 | 105 | 210 |
| | Quartz sand | - | 680 | 0.42 | - | 3.36 | 196 | 285.6 |
| | Quartz sand | - | 860 | 0.41 | - | 3.63 | 256 | 361.2 |
| Kimura (2018) | Glass bead | 105-125 | 115 | 0.366 | - | 4.09 | 8.8 | 48.3 |

| | | | | | | | |
|---|---|---|---|---|---|---|---|
| Glass bead | 125-149 | 136 | 0.364 | - | 4.20 | 10.7 | 57.1 |
| Glass bead | 149-177 | 162 | 0.363 | - | 4.13 | 18.3 | 68.0 |
| Glass bead | 177-210 | 193 | 0.364 | - | 4.04 | 26.7 | 81.1 |
| Glass bead | 210-250 | 229 | 0.362 | - | 4.20 | 33 | 96.2 |
| Glass bead | 250-297 | 273 | 0.358 | - | 4.17 | 51 | 114.7 |
| Glass bead | 297-350 | 324 | 0.358 | - | 4.15 | 67.4 | 136.1 |
| Glass bead | 350-420 | 386 | 0.356 | - | 4.36 | 102.1 | 162.1 |
| Glass bead | 420-500 | 459 | 0.358 | - | 4.30 | 134.3 | 192.8 |
| Glass bead | 500-590 | 545 | 0.36 | - | 4.06 | 246.2 | 228.9 |
| Glass bead | 590-710 | 648 | 0.358 | - | 4.18 | 299 | 272.2 |
| Glass bead | 710-840 | 771 | 0.357 | - | 4.29 | 510.4 | 323.8 |
| Glass bead | 840-1000 | 917 | 0.356 | - | 4.15 | 611.9 | 385.1 |
| Silica sand | 105-125 | 115 | 0.379 | - | 4.02 | 7 | 48.3 |
| Silica sand | 125-149 | 136 | 0.378 | - | 4.27 | 10.9 | 57.1 |
| Silica sand | 149-177 | 162 | 0.378 | - | 4.21 | 16.6 | 68.0 |
| Silica sand | 177-210 | 193 | 0.378 | - | 4.16 | 20 | 81.1 |
| Silica sand | 210-250 | 229 | 0.38 | - | 4.24 | 27.5 | 96.2 |
| Silica sand | 250-297 | 273 | 0.38 | - | 4.15 | 45.4 | 114.7 |
| Silica sand | 297-350 | 324 | 0.38 | - | 4.07 | 70.5 | 136.1 |
| Silica sand | 350-420 | 386 | 0.38 | - | 4.12 | 89.9 | 162.1 |
| Silica sand | 420-500 | 459 | 0.381 | - | 4.17 | 133.7 | 192.8 |
| Silica sand | 500-590 | 545 | 0.383 | - | 4.09 | 189.6 | 228.9 |
| Silica sand | 590-710 | 648 | 0.385 | - | 4.12 | 270.8 | 272.2 |
| Silica sand | 710-840 | 771 | 0.388 | - | 4.10 | 391.7 | 323.8 |
| Silica sand | 840-1000 | 917 | 0.389 | - | 3.95 | 558.6 | 385.1 |
| Fujikawa sand | 149-177 | 162 | 0.442 | - | 3.75 | 14.4 | 68.0 |
| Fujikawa sand | 210-250 | 229 | 0.421 | - | 3.83 | 27.8 | 96.2 |
| Fujikawa sand | 250-297 | 273 | 0.419 | - | 3.79 | 42.9 | 114.7 |
| Fujikawa sand | 297-350 | 324 | 0.416 | - | 3.88 | 56.5 | 136.1 |
| Fujikawa sand | 350-420 | 386 | 0.413 | - | 3.90 | 81.8 | 162.1 |
| Fujikawa sand | 420-500 | 459 | 0.414 | - | 3.93 | 123.8 | 192.8 |
| Fujikawa sand | 500-590 | 545 | 0.415 | - | 3.92 | 176.8 | 228.9 |
| Fujikawa sand | 590-710 | 648 | 0.415 | - | 3.91 | 234.6 | 272.2 |

$\overline{D}$: average grain diameter, $\phi$: porosity, $m$: cementation exponent in Archie's law (Archie, 1942), $F$: formation factor, $k$: permeability, $d_c$: critical pore diameter

**Figure captions**

Figure 1. Estimated permeability via (a) the Kozeny-Carman model, Eq. (1), (b) the Revil and Cathles (1999) model, Eq. (3), (c) the RGPZ model, Eq. (4), and (d) critical path analysis, Eq. (5), with $c$ = 72.2, versus measured permeability for 105 uniform glass bead and sand packs from the literature. Salient properties of each pack are presented in Table 1. The red dashed line represents the 1:1 line.

Figure 2. Measured permeability, $k$, against (a) average grain diameter, $\overline{D}$, and (b) critical pore diameter, $d_c$, for 105 uniform glass bead and sand packs studied here. Salient properties of each pack are presented in Table 1.

Figure 3. Measured formation factor, $F$, against measured porosity, $\phi$, for 78 uniform glass bead and sand packs studied here. The red solid line represents $F = \phi^{-1.5}$ (Sen et al., 1981), theoretically derived for mono-sized sphere packs using the self-consistent model. Salient properties of each pack are presented in Table 1.

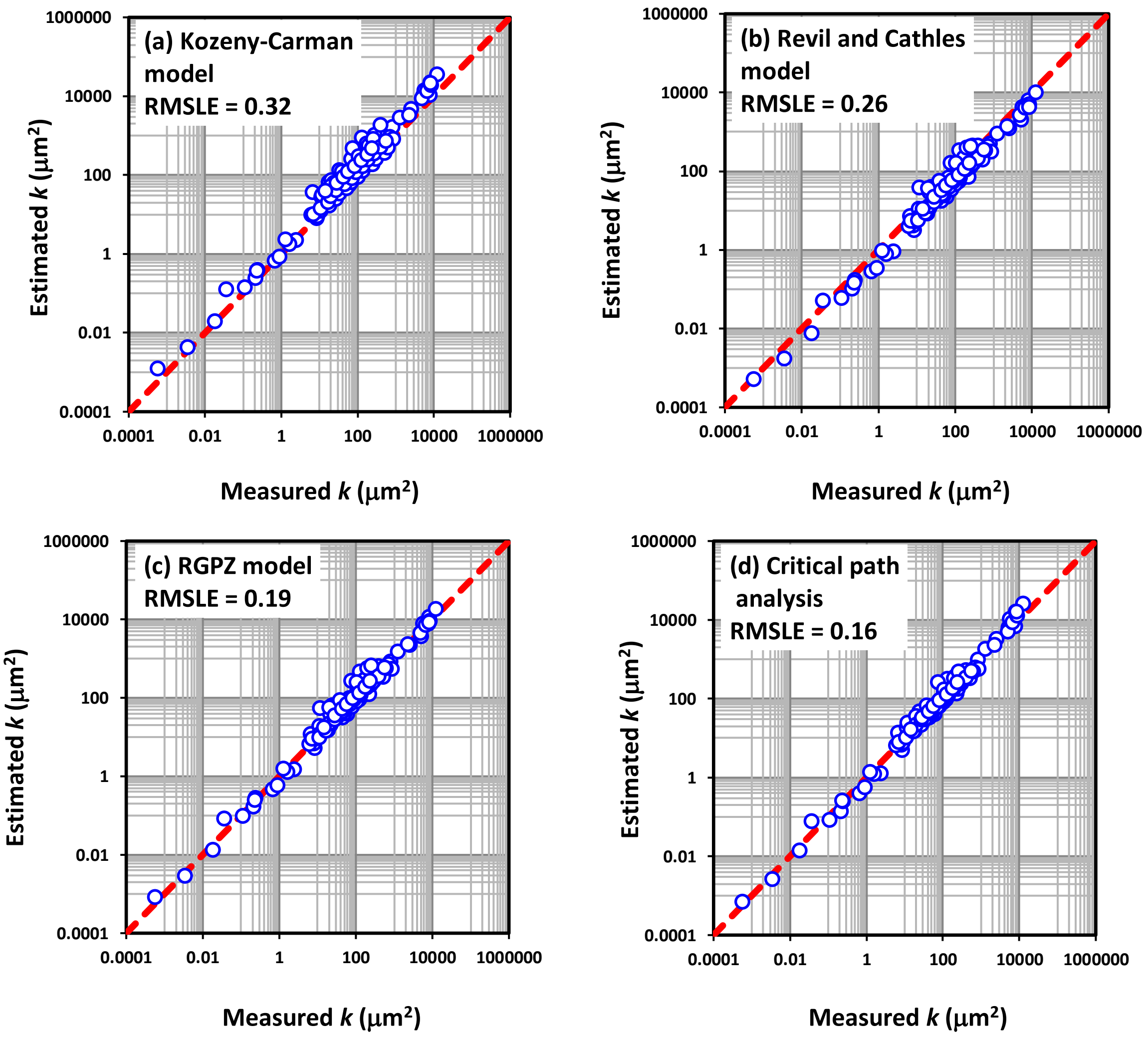

Fig. 1.

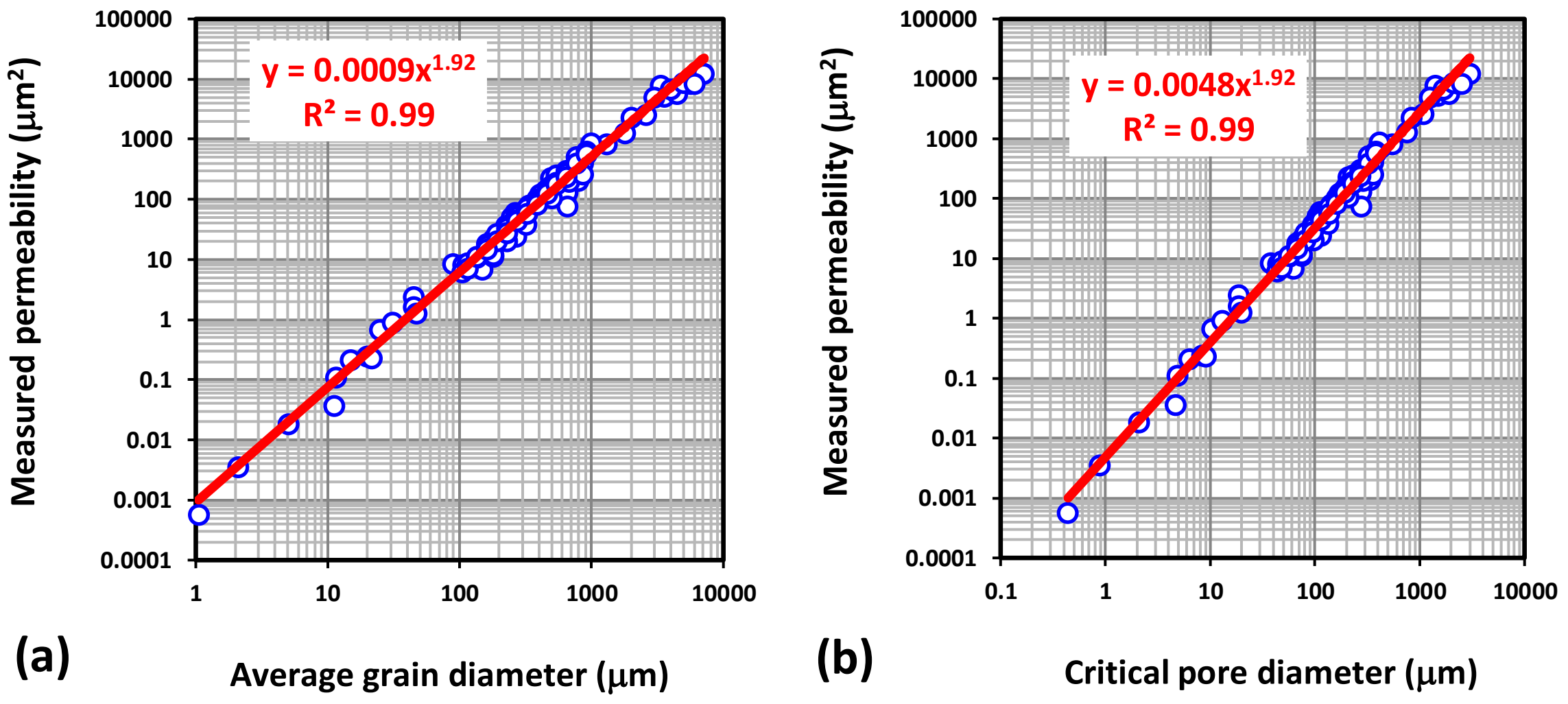

Fig. 2.

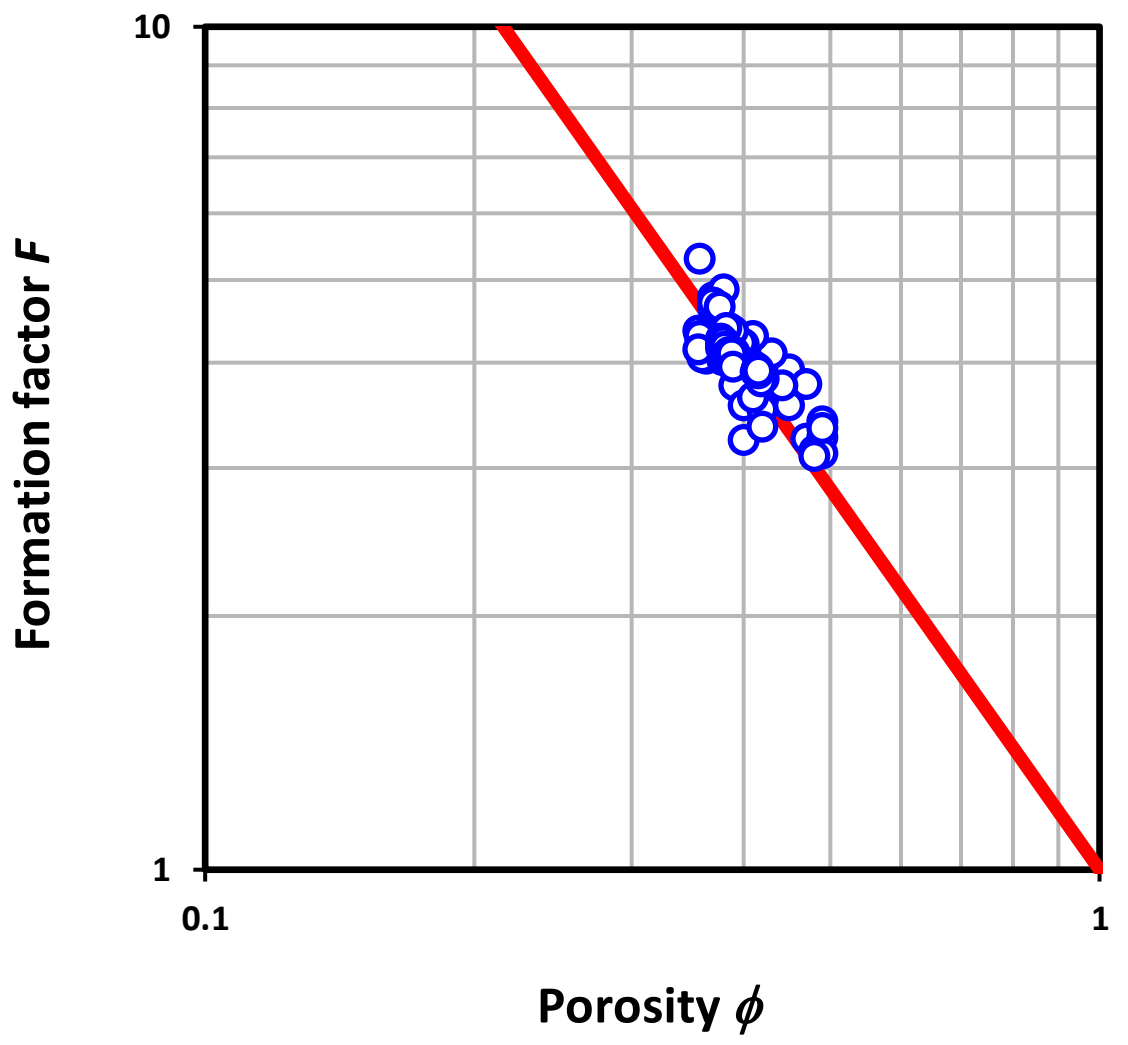

Fig. 3.